\documentclass[preprint,showkeys,preprintnumbers,amsmath,amssymb]{revtex4}

\usepackage{graphicx}
\usepackage{xcolor}
\usepackage{dcolumn}
\usepackage{bm}

\begin{document}

\author{Romeu Rossi Jr.}
\email{romeu.rossi@ufv.br}
\affiliation{Federal University of Vi{\c c}osa (UFV) - Campus Florestal,
LMG818 Km6, Minas Gerais, Florestal, Brazil}

\author{Leonardo A. M. Souza}
\email{leonardoamsouza@ufv.br}
\affiliation{Federal University of Vi{\c c}osa (UFV) - Campus Florestal,
LMG818 Km6, Minas Gerais, Florestal, Brazil}

\title{Causal Emergence in Quantum Mechanics}

\begin{abstract}
Causal emergence is brought about when a coarse-grained description of a physical system is more effective (more deterministic and/or less degenerate) than the fine-grained corresponding model. We show, for the first time to our knowledge, a causal emergence in a quantum system: a atomic Mach-Zehnder interferometer with two which-path detectors. The atomic wave-like or particle-like descriptions are related, respectively, to \emph{coarse-grained} and \emph{fine-grained} models. The predictions of the atomic position after the passage through the interferometer when the to \emph{coarse-grained} description is considered are more effective than the corresponding \emph{fine-grained} model. We conclude that quantum eraser measurements yields the causal emergence in this system. 
\end{abstract}

\keywords{Causality; Causal Emergence}
\maketitle

Generally it is assumed that once a fine-grained, detailed and exhaustive description of a physical system is formulated, all the features of a corresponding coarse-grained model are also established. This assumption is based on the idea that fine-grained descriptions can capture the essence of the phenomena, whereas by coarse-graining the result would leat to a lost of predictive and descriptive power. In some cases, when a fine-grained description is not yet available, a coarse-grained model is employed for practical purpose. Nevertheless, one can consider that the aim of scientific research is to unveil the detailed structure of the phenomena through fine-grained description.

Recently E.P. Hoel et. al. \cite{pnas} have shown that this reductionist approach can not be considered as a general feature of scientific explanation. They have introduced a quantity, so called Effective Information (EI) \cite{defEI}, that measure \emph{the effectiveness degree of a theoretical model}, and also they showed some examples in which a coarse-grained description is more effective (more deterministic and/or less degenerate) than the fine-grained corresponding model\cite{com}. The authors call this feature ``Causal Emergence''. 

In the mentioned work \cite{pnas}, causal models are employed in order to make the comparison between fine and coarse-grained descriptions. A causal explanation of a phenomena can be formulated with different levels of details, in relation to the number of variables used in the causal description, or to the time scale considered in the evolution of the system. A detailed causal model, associated to a fine-grained description of a system, determines the structure of the causal model of the corresponding less detailed model, that is constructed by coarse-graining the first structure. However, as shown in \cite{pnas,ent}, the EI calculated to the coarse-grained description can exceed the value of EI for the corresponding fine-grained causal model. Their conclusion implies, at least for the examples studied in \cite{pnas}, that the effectiveness of the theoretical model of the coarse-grained description can be greater than the fine-grained.

The systematic approach to causality given by the theory of causal Bayesian network \cite{pearl} made this a topic of increasing interest among physicists. The possibility of applying causal models in quantum systems is a current subject of debate in the literature, examples of incompatibility between causal models and quantum systems are shown in References \cite{rossi, wood}, and in Reference \cite{chaves} it was shown a quantum system that is compatible with causal description. Alternative approaches that aim to formulate causal models more suitable for quantum systems have also been considered in the Literature \cite{art1, art2, art3, art4, art5, art6, art7}.

In the study of causal emergence, a causal model can, as mentioned explicitly here, ``represent state transitions, like Markov chains, or may represent the influence or connectivity of elements, such as circuit diagrams, directed graphs (also called causal Bayesian networks), networks of interconnected mechanisms, or neuron diagrams''\cite{ent}. These causal models are well suitable to describe classical systems, and rigorous investigations on the adequacy of the use of causal models in the representation of the present quantum system is not part of the scope of this work. Here, we consider a classical causal model (Markov chain) in which the states are composed by classical variables whose values are determined by the results of measurements in the quantum system. The value of these variables in the initial state are defined by the quantum state preparation. The transition probabilities, which are indicated in the Markov chains, are given by quantum evolution of the initial state. It is important to notice that the  Markov chains considered here are not quantum Markov chains.

Here we show, for the first time to our knowledge, a causal emergence yield by quantum features. We consider a Mach Zehnder interferometer with two which-path detector and show that probabilities corresponding to atomic wave-like or particle-like behaviour constitute a case of causal emergence. 

We consider two causal models, a fine-grained and a coarse-grained, for the same experimental setup. Each model responds differently for the classical question: ``what is the trajectory of a photon in a Mach-Zehnder interferometer?'' In the fist model, three variables are considered and they completely reveal the which-path information. We call this the fine-grained description. In the second model, only two variables are considered and the which-path information is not completely given, and we call this description the coarse-grained one. We show that a genuine quantum effect, the quantum eraser, can bring about  causal emergency. In this system the coarse-grained description is more effective (more deterministic and/or less degenerate) than the fine-grained with respect to the prediction of the atomic position after the passage through the interferometer.

This contribution is organized as follow: in section \ref{causal_section} we introduce the notions of causal emergence that we will use throughout the paper; in section \ref{quantum_system} we detail the physical system we are interested; in sections \ref{fine_sec} and \ref{coarse_sec} we show our results explicitly, for the fine and coarse-grained description of our model, respectively. Finally, in section \ref{conclusions} we conclude our work, and give some perspectives within this approach.
  
\section{Causal Emergence}\label{causal_section}

To characterize a causal emergence, causal models that corresponds to fine-grained and coarse-grained descriptions need to be considered. A quantitative analysis of the causal effectiveness of each model must be performed and a causal emergence occurs when the effectiveness calculated to the coarse-grained model is greater than the effectiveness of the fine-grained model. 

In the quantitative analysis we considered a system $S$, composed by classical random variables that are relevant to the description. A particular state of $S$ is represented by $s_{i}$. To measure causal effectiveness, the statistical relations between the initial state ($s_{0}$) and the state at a later instant ($s_{F}$) are studied. The initial state is set by interventions represented by the operator $do(s_{0})$ (defined in \cite{pearl}). 

In Ref. \cite{pnas}, state-dependent and state-independent investigations are given. In the first case, the quantity used to measure causal effectiveness is the ``effect information'' ($Ei$) that is defined as:
\begin{equation}
Ei(s_{0})= D_{KL}(p(s_{F}|do(s_{0}))||p(s_{F}))=\sum_{s_{F}}p(s_{F}|do(s_{0}))\log_{2}\frac{p(s_{F}|do(s_{0}))}{p(s_{F})},
\end{equation}
where $D_{KL}$ is the Kullback-Leibler divergence \cite{kull} that measures the difference between two probability distributions, in the present case between the constrained and unconstrained probabilities of $s_{F}$. It has been proven that $Ei(s_{0})$ can be written as a function of the ``determinism coefficient'' and the ``degeneracy coefficient'' (proof and the definitions of the coefficients are detailed in Reference \cite{pnas}), which correspond respectively to: (i) determinism coefficient $\rightarrow$ a measure of how deterministically the evolution of $s_{0}$ will bring the system to the state $s_{F}$; (ii) degeneracy coefficient $\rightarrow$ how exclusive is the state transition from $s_{0}$ to $s_{F}$. For a given model, higher values of $Ei(s_{0})$ indicate that the model is more deterministic and less degenerated.

In the state-independent analysis\cite{pnas}, the quantity called ``effective information" ($EI$) is used. It is defined as the mean value of effect information over all initial states $s_{0}$:
\begin{equation}
EI=\sum_{s_{0}}p(do(s_{0}))D_{KL}(p(s_{F}|do(s_{0}))||p(s_{F}))=\sum_{s_{0},s_{F}}p(do(s_{0}))p(s_{F}|do(s_{0}))\log_{2}\frac{p(s_{F}|do(s_{0}))}{p(s_{F})},
\end{equation}

In this work we have used the quantities mentioned in this section in order to study the physical model detailed in the next section.
  
\section{The Quantum System}\label{quantum_system}

\begin{figure}[h]
\centering
  \includegraphics[scale=0.4]{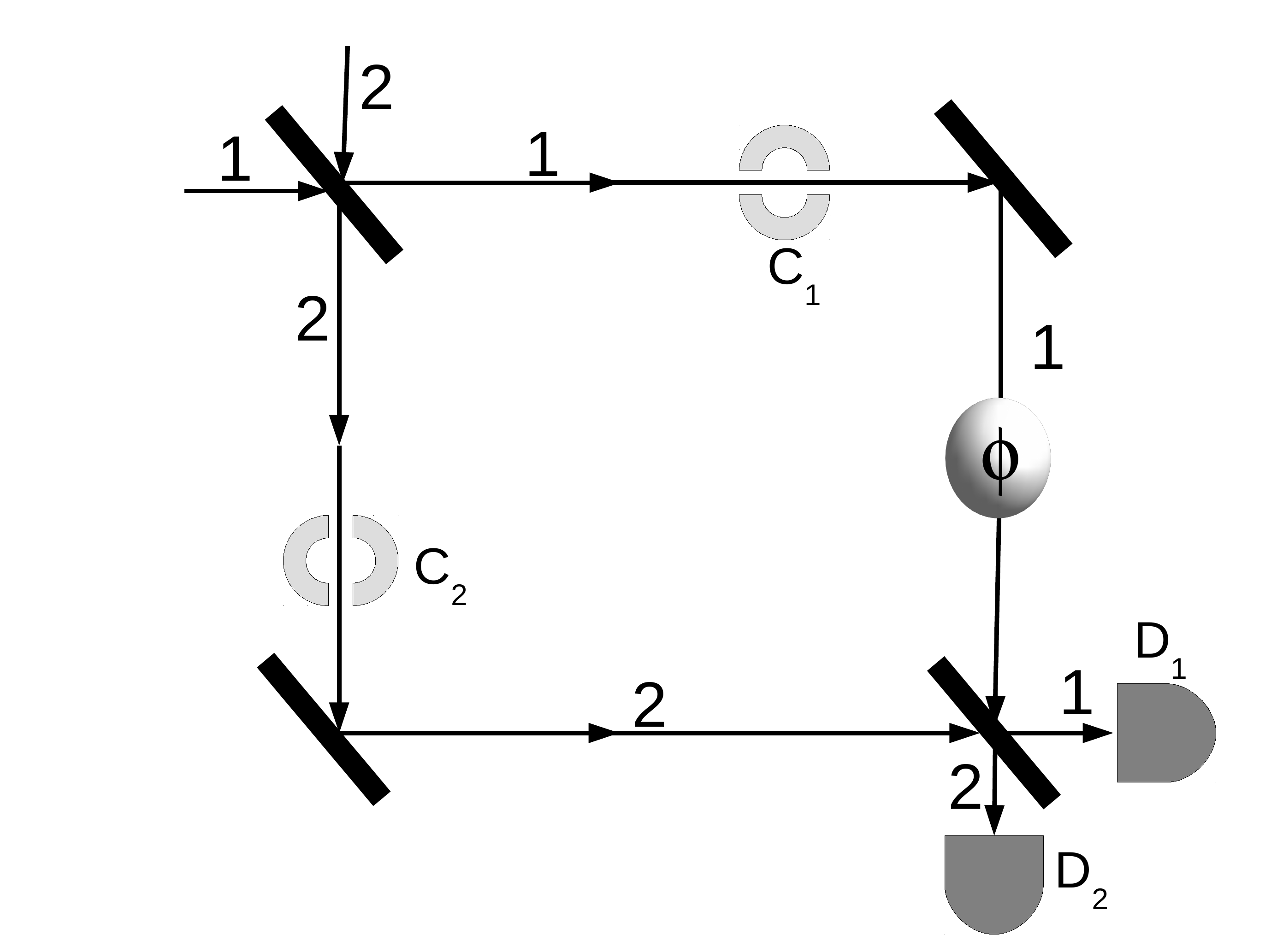}\\
  \caption{Detailed system. A Mach-Zehnder interferometer with a which-path detector ($C_1$ and $C_2$) in each arm.}\label{fig1}
\end{figure}

The quantum system we study here is composed by a two level atom that goes through a Mach-Zehnder interferometer with two which-path detectors, as it is shown in Figure \ref{fig1}. The detectors are two QED cavities ($C_1$ and $C_2$), resonant with the atomic transition, as in the first quantum eraser gedanken experiment \cite{scully}. The cavities are prepared in the vacuum state and the atom in the excited state, while the interaction time is chosen such that the atom makes a transition to the ground level when passing through the cavity. Atomic which-path information is available in the entangle state after the interaction with the cavities. The atoms are detected in each exit of the second beam splitter by the atomic detector $D_1$ and $D_2$. The detection probabilities (in $D_1$ and $D_2$) characterizes the wave-like or particle-like phenomena: they depend on the measurements performed in the cavities subsystem. Following the definitions given in Ref. \cite{bergou}, two types of measurement are of particular interest: which-alternative measurement and the quantum-erasure measurement. The first aims to reveal information about the atomic path, therefore it yield atomic detection probabilities that are associated with particles-like behavior. The latter aims to increase the visibility in the atomic subsystem, it produces atomic detection probabilities that are associated with wave-like behavior.

A detailed description can be written as follows: the source S emits two level atoms (prepared initially in the excited state $|e_{a}\rangle$) into the Mach-Zehnder interferometer, along the direction denoted by $|1_{a}\rangle$ or $|2_{a}\rangle$, with equal probability, as shown in Figure \ref{fig1}. The interaction time between the atom and each cavity correspond to a $\pi$ pulse, therefore, after the interaction with the cavity modes the atomic excitation is transferred to the cavities subsystem. The photon number in each cavity mode reveal the atomic which-path information, where the state $|1_{C_1},0_{C_2}\rangle$ ($|0_{C_1},1_{C_2}\rangle$) is associated with the atomic path $|1_{a}\rangle$ ($|2_{a}\rangle$). 

When the system is prepared in the initial state $|\psi(0)\rangle = |e_{a}\rangle|1_{a},0_{C_1},0_{C_2}\rangle$ the state of the global system after the second beam splitter and before the atomic detection is:
\begin{equation}
|\psi(t)\rangle_{1} = \frac{1}{2}\left[-|1_{a}\rangle\left(e^{i\phi}|1_{C_1},0_{C_2}\rangle+|0_{C_1},1_{C_2}\rangle\right)+i|2_{a}\rangle\left( e^{i\phi}|1_{C_1},0_{C_2}\rangle-|0_{C_1},1_{C_2}\rangle\right)\right]\label{evo1}.
\end{equation} On the other hand, when the system is prepared in the initial state $|\psi(0)\rangle = |e_{a}\rangle|2_{a},0_{C_1},0_{C_2}\rangle$, the state (before atomic detection) is:
\begin{equation}
|\psi(t)\rangle_{2} = \frac{1}{2}\left[i|1_{a}\rangle\left(-e^{i\phi}|1_{C_1},0_{C_2}\rangle+|0_{C_1},1_{C_2}\rangle\right)-|2_{a}\rangle\left( e^{i\phi}|1_{C_1},0_{C_2}\rangle + |0_{C_1},1_{C_2}\rangle\right)\right]\label{evo2},
\end{equation} We consider $\phi=0$ and omit the atomic excitation degree of freedom in \eqref{evo1} and \eqref{evo2} since this term is factored throughout the evolution.

After the interaction with the atom, the excitation is transferred to the cavities subsystem, that were initially in the vacuum  state. The subsystem $(C_1,C_2)$ is then restricted to the one excitation subspace $\{|1_{C_1},0_{C_2}\rangle,|0_{C_1},1_{C_2}\rangle\}$, and Pauli matrices can be defined for this two level subsystem. We consider, in the cavities subsystem, the observable: 
\begin{equation}
\hat{\sigma}=\vec{n}\cdot \vec{\sigma},
\end{equation}
where $\vec{n}=\left(\sin2\theta ~ \cos2\theta,~ \sin2\theta~\sin2\theta,~\cos2\theta\right)$ and
$\vec{\sigma}=\left(\sigma_{x},~\sigma_{y},~\sigma_{z}\right)$ are the Pauli spin operators within the two cavities subsystem. The eigenvectors and eigenvalues of $\hat{\sigma}$ are:
\begin{eqnarray}
\hat{\sigma}|M_{+}\rangle &=& |M_{+}\rangle\\
\hat{\sigma}|M_{-}\rangle &=& -|M_{-}\rangle,
\end{eqnarray} 
where $|M_{+}\rangle =\alpha |1_{C_1},0_{C_2}\rangle + e^{i\gamma} \beta |0_{C_1},1_{C_2}\rangle $, $|M_{-}\rangle =\beta |1_{C_1},0_{C_2}\rangle - e^{i\gamma} \alpha |0_{C_1},1_{C_2}\rangle $, with $\alpha = \cos \theta$ and $\beta =\sin \theta$.

In the cavities subsystem, which-alternative and the quantum-erasure measurements may be performed. Formally, the quantity $K(\sigma)$, defined in Ref. \cite{bergou}, is a quantitative measure of what one can learn about the which-way information from a measurement of the observable $\sigma$. For the present system:
\begin{equation}
K(\sigma)= |\cos 2\theta|.
\end{equation}

A which-alternative measurement corresponds to $\theta=0, ~\pi/2$, when the maximum which-way information is revealed. When $\theta=\pi/4$, a measurement of $\sigma$ gives no information about the alternative of the atomic ``trajectory'' (where by trajectory we the mean arms $1$ or $2$), and this is called a quantum-erasure measurement. 

\section{Fine-Grained Description}\label{fine_sec}

First we analyse a description that reveals, at each step of the evolution in the interferometer of Figure \ref{fig1}, which path the atom takes. For this purpose, one must consider the case of which-alternative measurement, where $\theta=0, ~\pi/2$. Therefore, the observable is $\hat{\sigma}=\hat{\sigma_{z}}$, and the eigenstates that correspond to the measurement results $1$ and $-1$ are, respectively, $|1_{C_1},0_{C_2}\rangle$ and $|0_{C_1},1_{C_2}\rangle$. After a measurement of the observable $\hat{\sigma}=\hat{\sigma_{z}}$, the number of excitations in $C_1$ and $C_2$, and consequently the atomic path, are known. For both results ($1$ or $-1$) the atomic detection probabilities in $D_1$ or $D_2$ are $P(D_1)=P(D_2)=0.5$. We will show that, for this case, such uncertainty about the detection possibilities imply a zero value for the Effective Information (EI).

Formally, we consider that, after the measurement in the cavities, each result is associated with a binary classical variable $\mathbf{c_1}=\{0,1\}$ and $\mathbf{c_2}=\{0,1\}$ which express the value of the number of excitations measured in each cavity. A detection of an excitation in one cavity, which corresponds to $\mathbf{c_1}=1$ or $\mathbf{c_2}=1$, imply the revelation of the path the atom takes after the first beam splitter. This description, that considers $\mathbf{c_1}$ and $\mathbf{c_2}$, is here classified as \emph{fine-grained}, because it holds two variables that allows the knowledge of the atomic path (between the first and second beam splitter), in opposition to the coarse-grained description (considered later in this work), that holds a number of variables that is insufficient to reveal such knowledge.

We also define a classical variable $\mathbf{a}=\{1,2\}$ that is associated to the atomic position in the preparation stage and after the measurement: the first one corresponds to the input direction $1$ or $2$ shown in Figure \ref{fig1}; the latter corresponds to a detection in $D_1$ or $D_2$. The atomic excitation degree of freedom can be ignored in this analysis.

\begin{figure}[h]
  \includegraphics[scale=0.35]{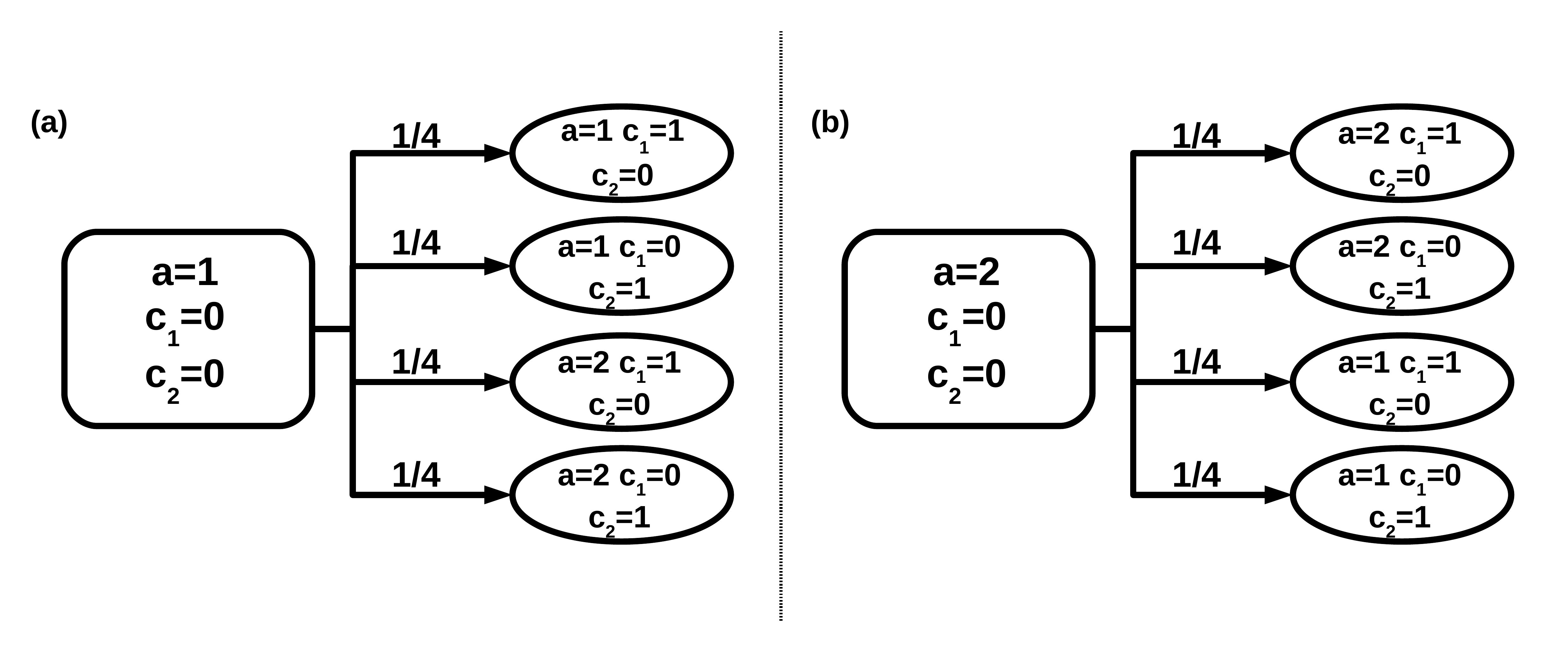}\\
  \caption{Markov chain associated with a \emph{fine-grained} description of our model.}\label{fig2}
\end{figure}

The possible initial states are: $s_{0}=\{(a(0)=1, c_1(0)=0, c_2(0)=0),(a(0)=2, c_1(0)=0, c_2(0)=0)\}$ and the possible final states are $s_{F}=\{(a(t)=1, c_1(t)=1, c_2(t)=0),(a(t)=1, c_1(t)=0, c_2(t)=1), (a(t)=2, c_1(t)=1, c_2(t)=0), (a(t)=2, c_1(t)=0, c_2(t)=1)\}$. The probability distribution of $s_{0}$ is $p(do(s_{0}))=0.5$, which correspond to the preparation of all initial states with the same probability (as required in Ref. \cite{pnas}). Each classical state of $s_{0}$ and $s_{F}$ is associated to a quantum states $|\psi(0)\rangle = |a(0), c_1(0), c_2(0)\rangle$ or $|\psi(t)\rangle = |a(t), c_1(t), c_2(t)\rangle$ where the values of $a$, $c_1$ and $c_2$ are equal to the corresponding values of the classical variables in the initial and final classical states. Notice that this complete correspondence between classical and quantum states are considered only after a measurement or a preparation of the system, during the evolution (in the time between a preparation and a measurement) the quantum states have no correspondence with classical states. The Markov chain associated is represented in Figure \ref{fig2}.

Quantum theory is used in the calculation of the conditional probability distribution. It gives the result $p(s_{F}|do(s_{0}))=|\langle \psi(t)|U|\psi(0)\rangle|^{2}=1/4$, which is equal (in this case) to the final state probability distribution $p(s_{F})=\sum_{s_{0}}p(s_{F}|do(s_{0}))p(do(s_{0}))=1/4$. Therefore, the \emph{fine-grained} description results:
\begin{equation}
Ei(s_{0})= \sum_{s_{F}}p(s_{F}|do(s_{0}))\log_{2}\frac{p(s_{F}|do(s_{0}))}{p(s_{F})}=0,
\end{equation} and consequently $EI=0$. This result should be compared with the \emph{coarse-grained} description, detailed hereafter.

\section{Coarse-Grained Description}\label{coarse_sec}

In this section we consider measurements of the observable $\hat{\sigma}$ with $0<\theta <\pi/2$. In this case, which-alternative information is not completely revealed, for instance, when $\theta=\pi/4$ any which-alternative information is completely removed (a quantum-erasure measurement), and the interference pattern can, therefore, be reconstructed. We show that the evolution of a quantum system in a Mach-Zehnder interferometer, when a quantum-erasure measurement is performed, corresponds to a model in which the value of $EI$ is maximum. In addition, we also show that this system is associated to a \emph{coarse-grained} description. These results, when compared to the ones shown in the last section, reveal causal emergence in a quantum system.

Formally, the \emph{coarse-grained} model is constructed from the substitution of the two variables $\mathbf{c_1}$ and $\mathbf{c_2}$ by a single variable $\mathbf{c}$. The classical variable $\mathbf{c}=\{0,1\}$ represents the results of a measurement of the excitation number in the cavities subsystem. Notice that when $\mathbf{c} =0$ both cavities are empty. However, when $\mathbf{c} =1$ one of the cavities has a single photon but it is not possible to identify in which cavity whatsoever. We also consider the variable $\mathbf{a}=\{1,2\}$, defined in the previous section, associated to the atomic position in the preparation stage and after the measurement.

The possible initial sates are: $s_{0}=\{(a(0)=1, c(0)=0),(a(0)=2, c(0)=0)\}$ and the possible final states are $s_{F}=\{(a(t)=1, c(t)=1)$ and $(a(t)=2, c(t)=1)$. As in the previous section, the probability distribution of $s_{0}$ is $p(do(s_{0}))=0.5$, which correspond to the preparation of all initial states with the same probability (as required in \cite{pnas}). Each classical state of $s_{0}$ and $s_{F}$ is associated to a quantum states $|\psi(0)\rangle = |a(0), c(0)\rangle$ or $|\psi(t)\rangle = |a(t), c(t)\rangle$ where the values of $a$ and $c$ are equal to the corresponding values of the classical variables in the initial and final classical states. 

As it is done in quantum eraser experiments and models \cite{scully, souza2013, walborn, Rueckner, Torres, ma}, let us analyze separately fringes and anti-fringes: if a measurement of $\hat{\sigma}$ gives the result $1$, fringes are obtained. When $\theta=\pi/4$, $\gamma=0$ and $\phi=0$ the conditional probabilities $p(s_{F}|do(s_{0}))=|\langle \psi(t)|U|\psi(0)\rangle|^{2}$ are:
\begin{eqnarray}
p(s_{F}=1,1|s_{0}=1,0)&=&1,\\
p(s_{F}=2,1|s_{0}=1,0)&=&0,\\
p(s_{F}=1,1|s_{0}=2,0)&=&0,\\
p(s_{F}=2,1|s_{0}=2,0)&=&1.
\end{eqnarray}

The values of these probabilities reflect the quantum eraser, and the interference pattern is reconstructed when the which-path information is erased. In the Mach-Zehnder interferometer shown in Figure \ref{fig1}, with equal optical path length ($\phi=0$), an atom sent in direction $a(0)=1$ ($a(0)=2$) will exit in the direction $a(t)=1$ $(a(t)=2)$,  characterizing the interference pattern of this system. The probabilities $p(s_{F})=\sum_{s_{0}}p(s_{F}|do(s_{0}))p(do(s_{0}))$ are:
\begin{eqnarray}
p(s_{F}=1,1)&=&0.5,\\
p(s_{F}=2,1)&=&0.5.
\end{eqnarray}

\begin{figure}[h]
\centering
  \includegraphics[scale=0.35]{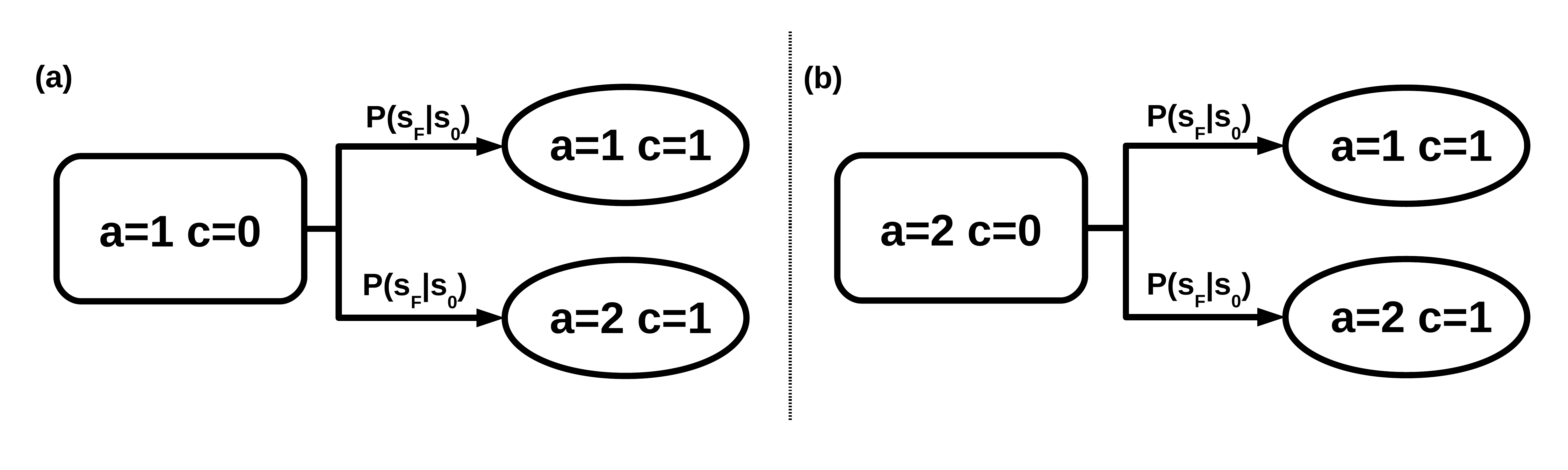}\\
  \caption{Markov chain associated with a \emph{coarse-grained} description of our model.}\label{fig3}
\end{figure}

Therefore, the Effect Information and the Effective Information are, respectively, $Ei(s_{0})=1$ and $EI=1$ for the \emph{coarse-grained} description. The values are greater than the values of $Ei$ and $EI$ calculated for the \emph{fine-grained} model, an explicity expression of causal emergence in the present quantum system. The wave-like behavior allow us to construct a more deterministic and/or less degenerate causal model (show in Figure \ref{fig3}), and this behavior is recovered by \emph{coarse-graining} the first description, or in other words by erasing which-path information. The quantum eraser phenomena plays a constitutive role in this case of causal emergence: it is responsible for the recovery of the wave-like behavior, and correspondingly to a coarse-grained description.      

We consider now a general case, with no specific values of $\theta$, and show that causal emergence also arises. We can associate a measurement of the operator $\hat{\sigma}$ with result $\{1\}(\{-1\})$ returning fringes (anti-fringes). Let us consider fringes and the initial state $s_{0}=(a(0)=1, c(0)=0)$. The conditional probabilities $p(s_{F}|do(s_{0}))=|\langle \psi(t)|U|\psi(0)\rangle|^{2}$ are:  
\begin{eqnarray}
p(s_{F}=1,1|s_{0}=1,0)&=&\left|\frac{-e^{i\phi}\alpha-e^{-i\gamma}\beta}{N_{1,+}}\right|^{2}, \\
p(s_{F}=2,1|s_{0}=1,0)&=&\left|\frac{e^{i\phi}\alpha-e^{-i\gamma}\beta}{N_{1,+}}\right|^{2}, \\ 
p(s_{F}=2,1|s_{0}=1,0)&=&\left|\frac{e^{i\phi}\alpha-e^{-i\gamma}\beta}{N_{1,+}}\right|^{2}, \\
p(s_{F}=2,1|s_{0}=2,0)&=&\left|\frac{-e^{i\phi}\alpha - e^{-i\gamma}\beta}{N_{2,+}}\right|^{2}
\end{eqnarray} where $\alpha = \cos \theta, \beta = \sin \theta$, and $N_{1,+}, N_{2,+}$ are normalization factors due to measurement processes.

\begin{figure}[h]
\centering
  \includegraphics[scale=1]{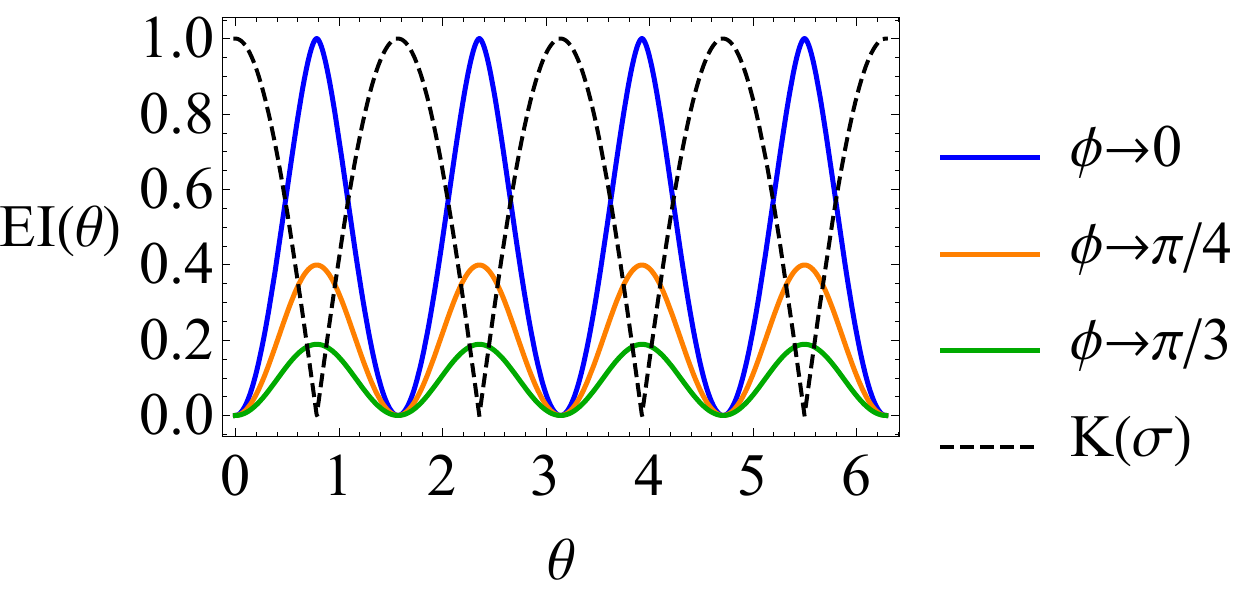}\\
  \caption{Effective Information (EI) in function of $\theta$ for different values of $\phi$. If $\theta$ is a multiple of $\pi/4(\pi/2)$, the EI aquires its maximum(minimum) value.}\label{fig4}
\end{figure}

The probabilities for the final states are calculated as $p(s_{F})=\sum_{s_{0}}p(s_{F}|do(s_{0}))p(do(s_{0}))$. The Effect Information for the initial state $s_{0}=(a(0)=1, c(0)=0)$ is:
\begin{equation}
Ei(s_{0})= \sum_{s_{F}}p(s_{F}|s_{0})\log_{2}\frac{p(s_{F}|s_{0})}{p(s_{F})},
\end{equation} and is plotted in Figure \ref{fig4}, in function of $\theta$. In previews sections we study the cases with $\theta =\pi/2$ (\emph{fine-grained} description) and $\pi/4$ (\emph{coarse-grained} description), this cases are associated respectively to the maximum and minimum value of $K(\sigma)$. Every intermediate value of $\theta$ corresponds also to a coarse-grained description, since the measurements of the observable $\sigma$ do not reveal complete which-path information in these cases, and one can not consider a classical causal model with $c_1$ and $c_2$ that describes the presence or absence of the atom in each path. The relation between $Ei(s_{0})$ and $K(\sigma)$ shows that when more which-path information is given (when $K(\sigma)$ increases) the value of  $Ei(s_{0})$ decreases.

\section{Conclusions}\label{conclusions}

Here we have presented for the first time (to our knowledge) a causal emergence yield by a quantum system. We consider a Mach-Zehnder interferometer with two which-path detector. Our results show that probabilities corresponding to atomic wave-like or particle-like behaviour can be related, respectively, to \emph{coarse-grained} and \emph{fine-grained} models. The which-way information is completely available in the \emph{fine-grained} and this makes the model less effective with respect to the prediction of the position of the atoms after the passage through the interferometer. For this model, the coarse-grained description is more effective (more deterministic and/or less degenerate) than the fine-grained. Our work paves way to further discussions related to causal emergency and fine/coarse-graining of multipartite systems in (closed and open) quantum systems, and how this concept can be tested.

\acknowledgments{The authors thanks Brazilian agencies CNPq and FAPEMIG for finantial support. LAMS also acknowledges INCT-IQ.}


\begin{thebibliography}{22}

\bibitem{pnas} E. P. Hoel, L. Albantakis, G. Tononi, \textit{Quantifying causal emergence shows that macro can beat micro}, Proceedings of the National Academy of Sciences \textbf{110}, 19790 (2013).

\bibitem{defEI} G. Tononi, O. Sporns, \textit{Measuring information integration}, BMC Neuroscience \textbf{4}, 31 (2003).

\bibitem{com} In \cite{pnas} the authors use the expressions ``micro-level'' or ``macro-level'' descriptions, but in the quantum context this could lead to confusion. Therefore, we use alternatively the nomenclature \emph{coarse-grained} or \emph{fine-grained} description. 

\bibitem{ent} E. P. Hoel, \textit{When the map is better than the territory}, Entropy \textbf{19}, 188 (2017).

\bibitem{pearl} J. Pearl, \textit{Causality: Models, Reasoning, and Inference.} 2nd ed. Cambridge: Cambridge University Press (2009).

\bibitem{rossi} R. Rossi, \textit{Restrictions for the causal inferences in an interferometric
system}, Physical Review A \textbf{96}, 012106 (2017).

\bibitem{wood} C. J. Wood, R. W. Spekkens, \textit{The lesson of causal discovery algorithms for quantum correlations: causal explanations of Bell-inequality violations require fine-tuning}, New Journal of
Physics \textbf{17}, 033002 (2015).

\bibitem{chaves} R. Chaves, G. B. Lemos, J. Pienaar, \textit{Causal Modeling the Delayed-Choice Experiment}, Physical Review Letters \textbf{120}, 190401 (2018).

\bibitem{art1} R. R. Tucci, \textit{Quantum Bayesian Nets}, International Journal of Modern Physics B \textbf{09}, 295 (1995). 

\bibitem{art2} M. S. Leifer, \textit{Quantum dynamics as an analog of conditional probability}, Physical Review A \textbf{74}, 042310 (2006).

\bibitem{art3} K. B. Laskey, \textit{Quantum Causal Networks}, arXiv:0710.1200 [quant-ph].

\bibitem{art4} M. S. Leifer, R. W. Spekkens, \textit{Towards a formulation of quantum theory as a causally neutral theory of Bayesian inference}, Physical Review A \textbf{88}, 052130 (2013).

\bibitem{art5} J. Henson, R. Lal, M. F. Pusey, \textit{Theory independent limits on correlations from generalized
Bayesian networks}, New Journal of Physics \textbf{16}, 113043 (2014).

\bibitem{art6} J. Pienaar, C. Brukner, \textit{A graph-separation theorem for quantum causal models}, New Journal of Physics \textbf{17}, 073020 (2015).

\bibitem{art7} F. Costa, S. Shrapnel, \textit{Quantum causal modelling}, New Journal of Physics \textbf{18}, 063032 (2016).

\bibitem{kull} S. Kullback, \textit{Information Theory and Statistics}, Dover Publications Inc.: Mineola, NY, USA, (1997).

\bibitem{scully} M. O. Scully, B. G. Englert, H. Walther, \textit{Quantum optical tests of complementarity}, Nature \textbf{351}, 111 (1991).

\bibitem{bergou} B. G. Englert, J. A. Bergou, \textit{Quantitative Quantum Erasure}, Optics Communications \textbf{179}, 337 (2000).

\bibitem{souza2013} R. Rossi, J. P. Souza, L. A. M. Souza, and M. C. Nemes, \textit{Multipartite quantum eraser in cavity QED}, Physical Review A \textbf{88}, 062102 (2013).

\bibitem{walborn} S. P. Walborn, M. O. Terra Cunha, S. P\'adua, and C. H. Monken, \textit{Double-slit quantum eraser}, Physical Review A \textbf{65}, 033818 (2002).

\bibitem{Rueckner} W. Rueckner and J. Peidle, \textit{Young's double-slit experiment with single photons and quantum eraser}, American Journal of Physics \textbf{81}, 951 (2013).

\bibitem{Torres} F. A. Torres-Ruiz, G. Lima, A. Delgado, S. P\'adua, and C. Saavedra, \textit{Decoherence in a double-slit quantum eraser}, Physical Review A \textbf{81}, 042104 (2010).

\bibitem{ma} X.-a. Ma, J. Kofler, and A. Zeilinger, \textit{Delayed-choice gedanken experiments and their realizations}, Reviews of Modern Physics \textit{88}, 015005 (2016).
 
\end{thebibliography}
\end{document}